\documentclass[12pt]{article}
 \input{epsf.sty}
 \input{epsf.tex}
 \newcommand{\be}{\begin{equation}}
 \newcommand{\ee}{\end{equation}}
 \newcommand{\ba}{\begin{eqnarray}}
 \newcommand{\ea}{\end{eqnarray}}
 \def\Journal#1#2#3#4{{#1} {#2}, #3 (#4)}

 \def\ASP{Astropart.\ Physics}
 \def\CQG{Class.\ Quantum Grav.}
 \def\GRG{Gen.\ Rel. Grav.}
 \def\IJMPA{Int.\ J.\ Mod.\ Phys.\ A}
 \def\IJMPD{Int.\ J.\ Mod.\ Phys.\ D}
 
 \def\MPLA{Mod.\ Phys.\ Lett.\ A}
 \def\NPB{Nucl.\ Phys.\ B}
 
 \def\PLB{Phys.\ Lett.\  B}
 \def\PRL{Phys.\ Rev.\ Lett.}
 \def\PRD{Phys.\ Rev.\ D}
 \def\PRP{Phys.\ Rept.}
 
 \def\laq{\,\raise 0.4ex\hbox{$<$}\kern -0.8em\lower 0.62ex\hbox{$\sim$}\,}
 \def\gaq{\,\raise 0.4ex\hbox{$>$}\kern -0.7em\lower 0.62ex\hbox{$\sim$}\,}

 \def\arccosh{\,{\rm arccosh}\,}
 
\begin{document}
\title{Pre-big bang in M-theory\thanks{This work is
supported in part by funds provided by the U.S.\ Department of Energy (D.O.E.)
under cooperative research agreement DE-FC02-94ER40818.}}
\author{Marco Cavagli\`a\footnote{Postal address:
Massachusetts Institute of Technology,
Marco Cavagli\`a 6-408, 77 Massachusetts Avenue,
Cambridge MA 02139-4307, USA. E-mail: cavaglia@mitlns.mit.edu}\\ 
Center for Theoretical Physics\\
Laboratory for Nuclear Science and Department of Physics,\\
Massachusetts Institute of Technology\\
77 Massachusetts Avenue, Cambridge MA 02139, USA\\
and\\
INFN, Sede di Presidenza, Roma, Italia}
\date{(MIT-CTP-3087, hep-th/0103095. \today)}
\maketitle
\begin{abstract}
We discuss a simple cosmological model derived from M-theory. Three assumptions
lead naturally to a pre-big bang scenario: (a) 11-dimensional supergravity
describes the low-energy world; (b) non-gravitational fields live on a
three-dimensional brane; and (c) asymptotically past triviality. 

\vskip 1em
\noindent
Pac(s) numbers: 98.80.Bp, 04.65.+e, 98.80.Hw, 04.50.+h
\end{abstract}
%
\section{Introduction and preliminaries}
Over the last two decades superstring theory \cite{strings} and, most recently,
M-theory \cite{M-th} have emerged as the most promising candidates for the
theory of quantum gravity. From a cosmological point of view, the key
theoretical question to be addressed is whether a successful inflationary model
can be constructed from superstring/M-theory. In superstring theories duality
relations between different regions of the moduli space provide a much richer
setting to investigate inflationary models than Einstein theory. Moreover, the
field content of superstring/M-theory in a given region of the moduli space is
determined unambigously on theoretical grounds. This provides a firm guidance
in the search for inflationary models which is absent in Einstein theory. On
the astrophysical/cosmological side a strong test for superstring/supergravity
theories is their compatibility with the standard model of early universe.
Observational cosmology becomes more and more efficient in constraining
cosmological parameters and the spectrum of  primordial perturbations.
Gravitational wave experiments may provide new and complementary constraints.
In the near future, cosmology will be the main laboratory to test whether
superstring/M-theory does really describe our universe or is just a nice
mathematical construction \cite{LWC}. In view of these developments
superstring/M-theory cosmology is a subject worth investigating.

In the cosmological setting, essential features of superstring/M-theory are the
presence of the dilaton, axion, Ramond-Ramond (R-R) forms and various moduli
fields, in addition to higher-curvature terms that appear in the low-energy
effective actions. The presence of these fields and string duality relations
have a profound impact on cosmological scenarios. For instance, simple
low-energy string models with graviton and dilaton lead to the so-called
pre-big bang (PRBB) scenario \cite{prebb} which is structurally different from
standard Einstein cosmology. In the PRBB scenario different branches of the
solution are related by time reflection and internal transformations --  the
scale factor duality \cite{duality}. The universe evolves from a weakly coupled
string vacuum state first to a radiation-dominated and then to a
matter-dominated Friedmann-Robertson-Walker (FRW) geometry through a region of
strong coupling and large curvature.

If M-theory is the ultimate theory of quantum gravity the low-energy world is
described by the 11-dimensional supergravity action \cite{M-th,Townsend}
\ba
S^{(11)}&=&{1\over l_{pl,11}^9}\int d^{11}x \sqrt{-g^{(11)}}\left[R^{(11)}(g^{(11)}_{ab})
-{1\over 48}F_{a_1\dots a_4}F^{a_1\dots a_4}+\right.\nonumber\\
&&\qquad\qquad\left. -{1\over 12^4\sqrt{-g^{(11)}}}\epsilon^{a_1 \dots a_3 
b_1 \dots b_4 c_1 \dots c_4}
A_{a_1 \dots a_3}F_{b_1 \dots b_4}F_{c_1 \dots c_4} \right]\,,
\label{I1}\ea
where $a_i,b_i,c_i=0\dots 10$, $F_{a_1 \dots a_4}=4\partial_{[a_1}A_{a_2 \dots
a_4]}$ is the 4-form field strength of the antisymmetric 3-form
potential $A_{a_1 \dots a_3}$, and $g^{(11)}$ denotes the determinant of the
11-dimensional metric $g^{(11)}_{ab}$. Equation (\ref{I1}) describes the
low-energy limit of M-theory. (Here and throughout the paper we use natural
units and set $l_{pl,11}$ such that the four-dimensional Planck length is
$l_{pl,4}=1$.)

Starting from equation (\ref{I1}) several different M-theory cosmological
models have been proposed in the literature. In particular, the idea of brane
world has emerged in the works of Lukas {\it et al} \cite{LOW} and Randall and
Sundrum \cite{RS}. According to this model our four-dimensional universe
emerges as the world volume of a 3-brane in a higher-dimensional spacetime. A
more standard approach \cite{cosm} deals with different classes of cosmological
solutions that reduce to solutions of string dilaton gravity. The analysis of
four-dimensional isotropic and homogeneous cosmologies derived from M-theory
and type IIA superstring theory (see last paper in \cite{cosm,CM}) has found
that form-fields associated with the Neveu/Schwarz-Neveu/Schwarz (NS-NS) and
R-R sectors play a different and crucial role in determining the dynamical
behavior of the solutions: the NS-NS fields, such as the axion, tend to forbid
inflation whereas the R-R fields have the opposite effect \cite{CM}.

The purpose of this paper is to investigate what cosmological scenario follows
from three simple assumptions:

\begin{itemize}

\item[(a)] M-theory is the correct description of nature;

\item[(b)] non-gravitational fields live on a three-dimensional brane propagating in
the 11-di\-men\-sio\-nal spacetime;

\item[(c)] the universe originated in the vacuum of the theory (asymptotically past
triviality, APT).

\end{itemize}

We will see that these simple postulates lead naturally to a PRBB cosmological
scenario, where two different branches of the solution are related by internal
symmetries of the model and the universe evolves from a weakly coupled string
vacuum state to a decelerated FRW geometry through a state with large
curvature. Although our model is probably too simple to give an accurate or
even acceptable description of our universe, we believe it is nevertheless
interesting and may represent a good starting point to discuss the PRBB
scenario in M-theory.
\section{The model}
Assumption (a) implies that the low-energy world is described by equation
(\ref{I1}). Following Witten \cite{M-th} we assume that the 11th dimension
is compactified on a circle $S_1$ of radius $R_{S_1}$. Carrying out a
Kaluza-Klein reduction we find
\ba
&&S^{(10)}={1\over l_{pl,10}^8}\int d^{10}x \sqrt{-g^{(s)}} \left[e^{-\Phi_{10}}
\left(R^{(10)}(g^{(s)}_{mn})
+\left(\nabla \Phi_{10} \right)^2-{1\over 12}H_{mnp}H^{mnp}
\right)\right.\nonumber\\
&&\qquad\qquad\left. -\frac{1}{48}F_{mnpq}F^{mnpq}-{1\over 384\sqrt{-g^{(11)}}}
\epsilon^{m_1 m_2 n_1 \dots n_4 p_1 \dots p_4}
B_{m_1 m_2}F_{n_1 \dots n_4}F_{p_1 \dots p_4}\right]\,,
\label{II1}
\ea
where we have rescaled the ten-dimensional metric as
$g_{ab}=R_{S_1}^{-1}g^{(s)}_{ab}$ ($a,b\not=10$) and we have defined the
dilaton by $R_{S_1}=e^{\Phi_{10}/3}$. $H_{mnp}$ and $F_{mnpq}$ are the field
strengths of the potentials $B_{np}$ and $A_{npq}$, respectively. Note that we
have ignored the 1-form potential that arises from the Kaluza-Klein reduction.
Equation (\ref{II1}) is the effective action for massless type IIA superstring.
The first line of equation (\ref{II1}) corresponds to the NS-NS sector and the
second line corresponds to the R-R sector.

Assumption (b) implies, for consistency, that the ten-dimensional
geometry must be of the form $M_4\times C_6$ and that the only non-trivial
components of the field strengths are those associated with $M_4$. $C_6$ is a
six-dimensional compact space which we assume to be a generic (Ricci flat)
Calabi-Yau space, or, for sake of simplicity, a six-dimensional torus. Upon
compactification on the six-dimensional internal space $C_6$ we find
($l_{pl,4}=1$)
\be 
S=\int d^4x\sqrt{-g}\left[e^{-\Phi_4}\left(R^{(4)}(g_{\mu\nu})
+\left(\nabla\Phi_4\right)^2-6\left(\nabla\beta\right)^2-{1\over 12} 
H_{\mu\nu\lambda}H^{\mu\nu\lambda}\right)-{1\over 48}e^{6\beta} 
F_{\mu\nu\lambda\kappa}F^{\mu\nu\lambda\kappa}\right]\,,
\label{II2}
\ee
where the radius of the internal space is $R_{C_6}=e^\beta$ and the
four-dimensional dilaton field is $\Phi_{4}=\Phi_{10}-6\beta$. The field
equation for the four-form $F^{\mu\nu\lambda\kappa}$ can be solved and the
3-form $H^{\mu\nu\lambda}$ can be dualized. The final result is
\be
S=\int d^4x\sqrt{-g}\left[e^{-\Phi_4}\left(R^{(4)}(g_{\mu\nu})+\left( 
\nabla\Phi_4\right)^2-6\left(\nabla\beta\right)^2-{1\over 2} 
e^{2\Phi_4}\left(\nabla\sigma \right)^2 \right) 
-{1\over 2}{\cal Q}^2 e^{-6\beta}\right]\,,
\label{II3}
\ee
where $\sigma$ is the pseudo-scalar axion field dual to the 3-form,
$H^{\mu\nu\lambda}=e^{\Phi_4}\epsilon^{\mu\nu\lambda\kappa}
\nabla_{\kappa}\sigma$, and $F^{\mu\nu\lambda\kappa}={\cal Q}e^{-6\beta}
\epsilon^{\mu\nu\lambda\kappa}$. Equation (\ref{II3}) describes the world as
seen by the four-dimensional observer and is our starting point to investigate
M-theory cosmology. 

Since we are interested in homogeneous and isotropic cosmologies we impose the
metric ansatz
\be
ds^2_{(4)}\equiv
g_{\mu\nu}(x)dx^\mu dx^\nu=-N^2(t)dt^2+a(t)^2d\Omega_{3k}\,,\qquad
N(t)>0
\label{II4}
\ee
where $d\Omega_{3k}$ is a maximally symmetric three-dimensional unit metric
with curvature $k=0,\pm 1$, respectively. Moreover, in the spirit of the APT
postulate, we assume that the four-dimensional metric (\ref{II4}) is spatially
flat, i.e., we set $k=0$\footnote{Note that the APT, in its general form
\cite{BDV}, admits more general initial states than spatially flat spacetime,
namely, generic perturbative solutions of the low-energy string action which
lead to gravitational instability.}. By substituting equation (\ref{II4}) in
equation (\ref{II3}) and requiring for consistency that the modulus field
$\beta$, the dilaton $\Phi_4$, and the axion $\sigma$ depend only on $t$, the
density action per comoving volume in the physical spacetime becomes
\be
S=\int
dt\,\left[{1\over\mu}\left(3\dot\alpha^2-\dot\phi^2
+6\dot\beta^2+{1\over 2}\dot\sigma^2e^{2(3\alpha+\phi)}\right)
-\mu{1\over 2}{\cal Q}^2e^{3\alpha-\phi-6\beta}\right]\,,
\label{II5}
\ee
where $\alpha(t)=\ln[a(t)]$, and we have defined the `shifted dilaton' field
$\phi=\Phi_4-3\alpha$ and the Lagrange multiplier $\mu(t)=Ne^{\phi}>0$.
Finally, equation (\ref{II5}) can be cast in the canonical form
\be
S=\int dt \left[\dot\alpha p_\alpha+\dot\phi p_\phi+\dot\beta
p_\beta+\dot\sigma p_\sigma-{\cal H}\right]\,,
\label{II6}
\ee
where the Hamiltonian is
\be
{\cal H}=\mu H\,,\qquad H={1\over 24}\left[2p^2_\alpha-6p^2_\phi+p^2_\beta
+12{\cal Q}^2e^{3\alpha-\phi-6\beta}
\left(1+{p^2_\sigma\over {\cal Q}^2}e^{-9\alpha-\phi+6\beta}\right)\right]\,.
\label{II7}
\ee
The total Hamiltonian ${\cal H}$ is proportional to the non-dynamical variable
$\mu$ which enforces the constraint $H=0$.

We now make a further assumption. We assume that, at least at the beginning of
the evolution, the NS-NS axion field is negligible w.r.t.\ the R-R 4-form
field, namely, we consider a volume in the four-dimensional spacetime which is
nearly devoid of axions. Quantitatively, we require
\be
|{\cal Q}/p_\sigma|>>1\,.
\label{II8}
\ee
(Note that $p_\sigma$ is constant.) The dynamical behaviour at the beginning of
the evolution is described by the solution \cite{CM}
\be
\begin{array}{lllll}
\alpha&=&\displaystyle \alpha_0-{1\over 4}\ln
\left[\cosh\left(\kappa{\cal T}\right)\right]
-\alpha_1{\cal T}\,,\qquad
p_\alpha&=&\displaystyle -{3\kappa\over 2}\tanh\left(\kappa{\cal T}\right)
-6\alpha_1\,,\\\\\
\phi&=&\displaystyle \phi_0-{1\over 4}\ln
\left[\cosh\left(\kappa{\cal T}\right)\right]-
\phi_1{\cal T}\,,\qquad
p_\phi&=&\displaystyle {\kappa\over 2}\tanh\left(\kappa{\cal T}\right)
+2\phi_1\,,\\\\
\beta&=&\displaystyle \beta_0+{1\over 4}\ln
\left[\cosh\left(\kappa{\cal T}\right)\right]-
\beta_1{\cal T}\,,\qquad
p_\beta&=&\displaystyle 3\kappa\tanh\left(\kappa{\cal T}
\right)-12\beta_1\,,
\end{array}\label{II9}
\ee
where 
\be
{\cal T}(t)=\int_{t_0}^t \mu(t') dt'\,,\qquad t>t_0\,.
\label{II10}
\ee
The constants of motion are related by (we choose $\kappa>0$ for simplicity)
\ba
\kappa^2+6\alpha_1^2-2\phi_1^2+12\beta_1^2-2H&=&0\,,
\label{II11}\\
3\alpha_0-\phi_0-6\beta_0-2\ln\left({\kappa\over |{\cal Q}|}\right)&=&0\,,
\label{II12}\\
3\alpha_1-6\beta_1-\phi_1&=&0\,.
\label{II13}
\ea
The dynamics of the model is determined essentially by $\alpha_1$. The constant
$\kappa$ determines the scale of the time evolution and can be reabsorbed in
the solution by defining the parameter $\tau=\kappa{\cal T}$ and the constants
$\xi=\alpha_1/\kappa$, $\chi=\phi_1/\kappa$, $\rho=\beta_1/\kappa$ and $Q={\cal
Q}/\kappa$. The constants $\alpha_0$, $\beta_0$ and $\phi_0$ are initial
conditions for the phase space coordinates $\alpha$, $\beta$ and $\phi$,
respectively.

The evolution proceeds monotonically in the gauge parameter $\tau$. The latter
is related to the proper time of the four-dimensional world by the relation 
\be
t_c(\tau)=\int_{\tau_0}^\tau d\tau'\, e^{-\phi(\tau')}\,,
\label{II14}
\ee
Since the integrand in Eq.\ (\ref{II14}) is positive defined, the time
evolution in the cosmic time $t_c$ follows the evolution in the gauge parameter
$\tau$.

The structure of the moduli space is conveniently described in the plane
($\xi,\rho$). The physical points are determined in this plane by
the two branches of the hyperbola (\ref{II11}) (see figure 1) 
$$
\hbox{$(+)$ branch: } \rho={3\over 5}\xi+{1\over 5}
\sqrt{4\xi^2+{5\over 12}}\,;\qquad
\hbox{$(-)$ branch: } \rho={3\over 5}\xi-{1\over 5}
\sqrt{4\xi^2+{5\over 12}}\,.
$$
\begin{figure}
\centerline{\epsfxsize=3.6in \epsffile{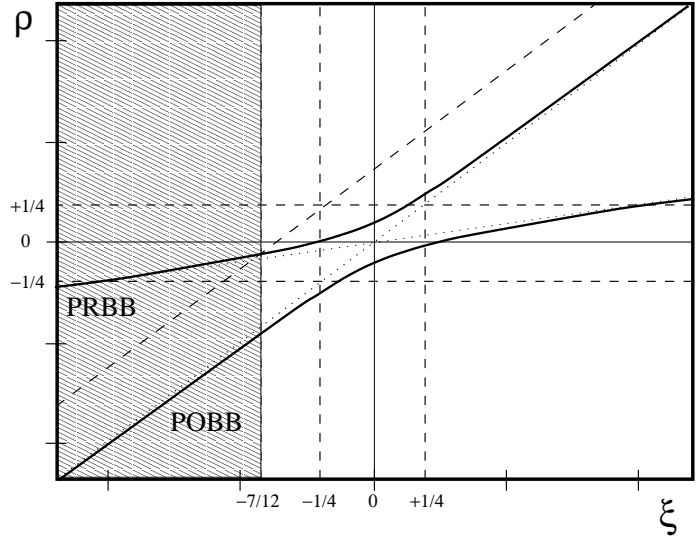}}
\caption{\small Parameter space for the R-R four-form dominated solution. The
physical points are represented by the two branches of the hyperbola
(\ref{II11}). The PRBB and POBB phases correspond to the portions of the
upper and lower hyperbola in the colored regions, respectively}
\label{figI}
\end{figure}
(The notation will be clearified soon.) The $(+)$ and $(-)$ branches are
characterized by a few distinctive kinematical and dynamical properties:  
\begin{center}
Table 1.\par
\begin{tabular}{|c||c||c|}
\hline
&{\bf $(+)$ branch}
&{\bf $(-)$ branch}\\ \hline
\hbox{\small 4D effective coupling $g$}
&\hbox{\small Increasing}
&\hbox{\small Decreasing}\\ \hline
\hbox{\small Initial 4D effective coupling}
&{\small Weak, perturbative}
&{\small Strong, non-perturbative }\\ \hline 
\hbox{\small 11D curvature $K$}
&{\small Increasing}
&{\small Decreasing }\\ \hline 
\hbox{\small Initial curvature scale}
&{\small Arbitrarily small}
&{\small Arbitrarily large}\\ \hline 
\end{tabular}
\bigskip
\end{center}
where $g\equiv e^\phi$ and $K\equiv
R^{(11)}{}_{\mu\nu\rho\sigma}R^{(11)}{}^{\mu\nu\rho\sigma}$. The $(+)$ and
$(-)$ branches remind one of the PRBB and post-big bang (POBB) branches in the PRBB
scenario \cite{prebb}. In the standard dilaton gravity models of PRBB
\cite{prebb} the two branches coincide with accelerated and decelerated scale
factors, respectively. Here, the presence of extra fields (the R-R 4-form)
makes the picture more complicated. For each of the two branches we find three
different dynamical behaviours of the external geometry according to the value
of $\xi$: expanding for $\xi<-1/4$ (accelerated for the $(+)$ branch and
decelerated for the $(-)$ branch), contracting for $\xi> 1/4$ (accelerated for
the $(+)$ branch and decelerated for the $(-)$ branch) and bouncing for
$-1/4<\xi<1/4$ (accelerated at early and late times for the $(+)$ branch and
decelerated for the $(-)$ branch)\footnote{We do not consider here the
`fine-tuned' cases $\xi=\pm 1/4$.}. The qualitative behaviour of the external
scale factor is represented in figure 2. 
\begin{figure}
\centerline{\epsfxsize=1.6in \epsffile{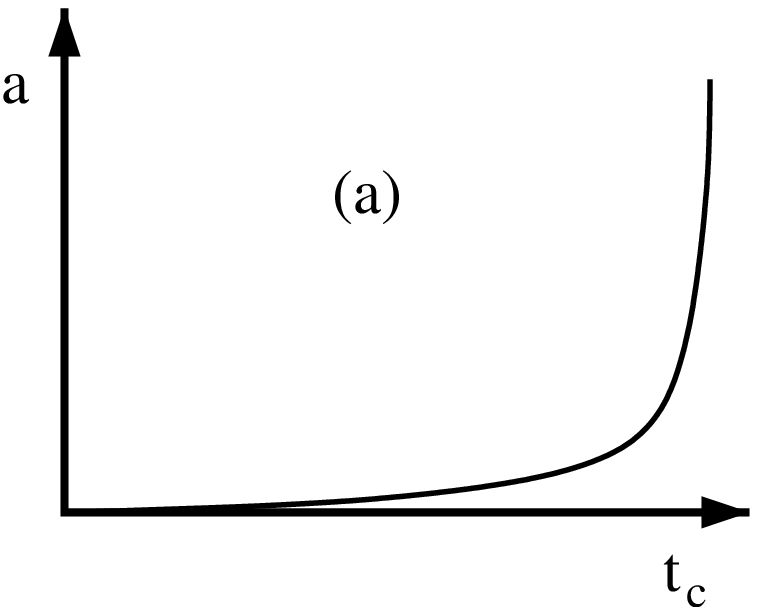}
\epsfxsize=1.6in\epsffile{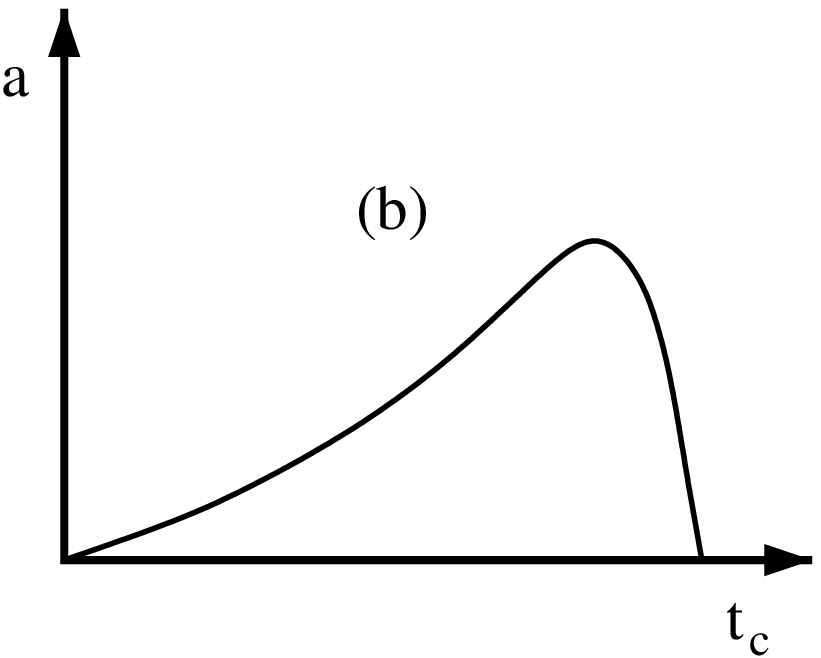}
\epsfxsize=1.6in\epsffile{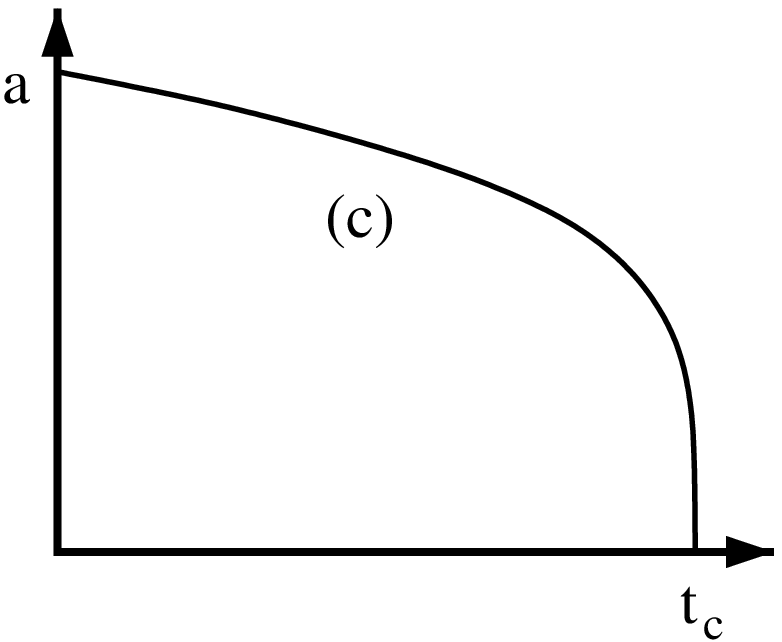}}
\centerline{\epsfxsize=1.6in \epsffile{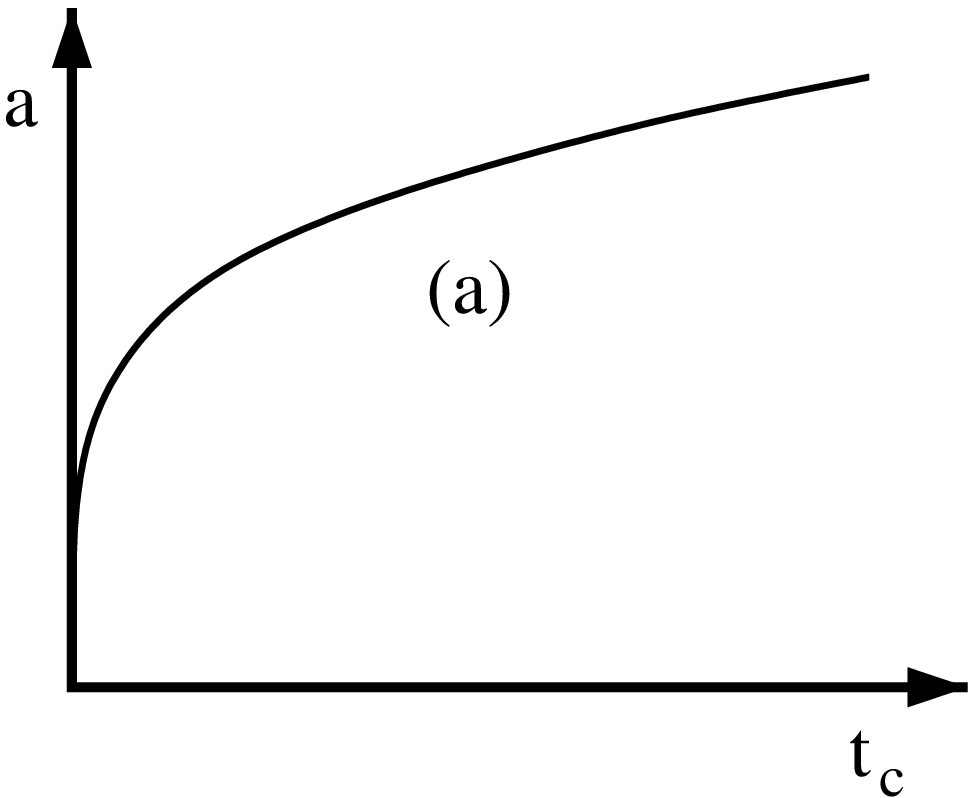}
\epsfxsize=1.6in\epsffile{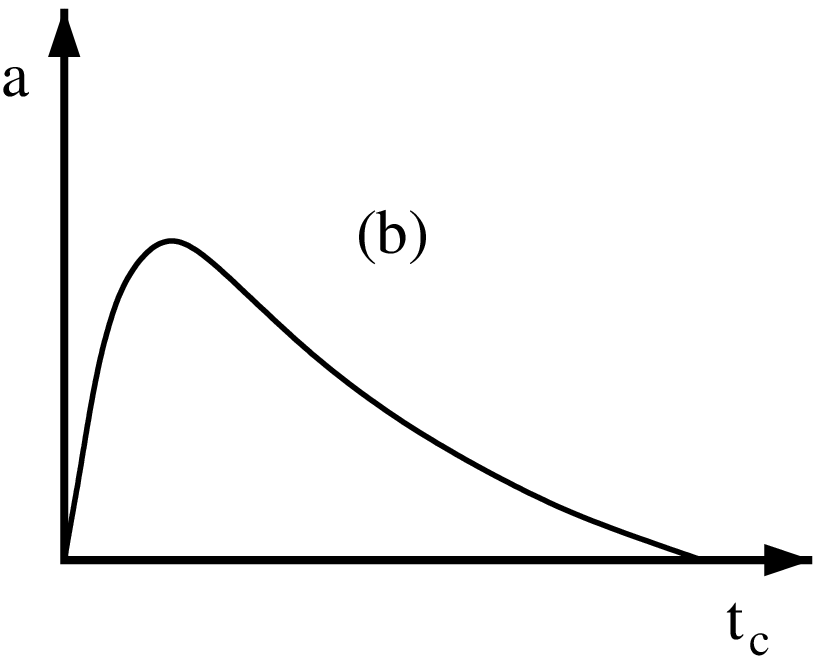}
\epsfxsize=1.6in\epsffile{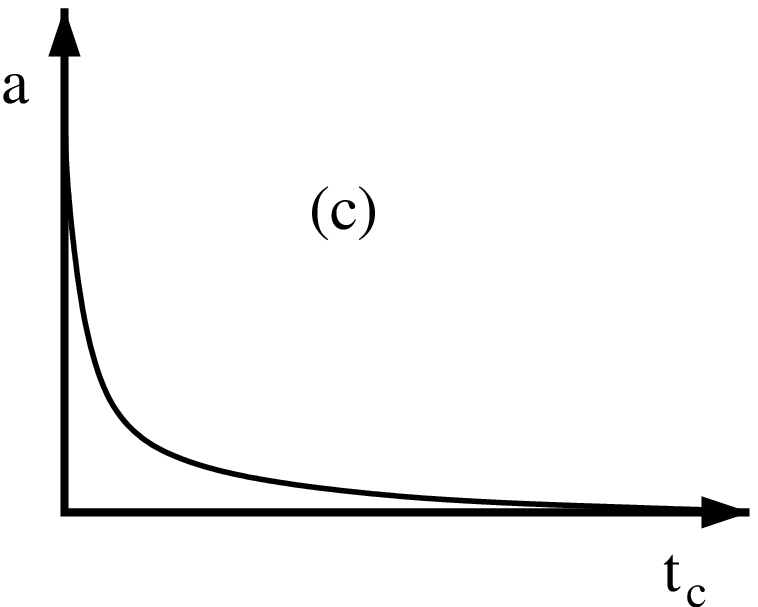}}
\caption{\small Qualitative behaviour of the external
scale factor for the $(+)$ branch (top figures) and $(-)$ branch (bottom
figures) for: (a) $\xi<-1/4$, (b) $-1/4<\xi<1/4$, and (c) $\xi>1/4$. Note
the symmetry $(+)\to(-)$ and $\xi\to -\xi$.}
\label{figII}
\end{figure}
The Hubble parameter and the deceleration parameter of the four-dimensional
world are
\be
H(t_c)=-{e^{\phi_0-\chi\tau}\over\cosh^{1/4}(\tau)}\left(\xi+{1\over 4}
\tanh(\tau)\right)\,,
\label{II15}
\ee
and
\be
q(t_c)=-{1\over \xi+\tanh(\tau)/4}\left(\xi+\chi+{1\over 2}
\tanh(\tau)-{1\over 4\cosh^{2}(\tau)(\xi+
\tanh(\tau)/4)}\right)\,,
\label{II16}
\ee
respectively. For $\tau\to\infty$ we have
\be
H(t_c)\approx-2^{1/4}e^{\phi_0}(\xi+1/4)e^{-\tau(\chi+1/4)}\,,
\label{II17}
\ee
and
\be
q(t_c)\approx -{1\over \xi+1/4}\left(\xi+\chi+1/2\right)\,.
\label{II18}
\ee
Note that the deceleration parameter is always finite.
\section{The scenario}
A physical description of our universe requires a large universe at late times.
This is achieved in the region of the moduli space $\xi<-1/4$. Since we are
assuming that at small times the 4-form potential is dominant with respect to
the axion potential term, the dynamics at the early stages of the evolution is
described by the solution above. The APT assumption (c) then implies that the
early universe must be described by the (+) branch solution of the hyperbola.
Indeed, for the $(-)$ branch of the hyperbola the 11-dimensional curvature
blows up at early times and approaches zero at large times. The converse is
true for the $(+)$ branch. The APT postulate requires that the evolution starts
in a low-energy state, i.e.\ in a state with small curvature. Clearly, the
lower branch does not satisfy this requirement. If we restrict attention to the
region $\xi<-1/4$ the similarity of the $(+)$ and $(-)$ branches to the PRBB
and POBB branches of the PRBB scenario is complete. Table 2 summarizes the
kinematical and dynamical properties of the plus and minus branch for
$\xi<-1/4$:  

\begin{center}
Table 2.\par
\begin{tabular}{|c||c||c|}
\hline
&{\bf $(+)$ branch}
&{\bf $(-)$ branch}\\ \hline
\hbox{\small Time evolution}
&\hbox{\small $\dot a>0$, $\ddot a>0$}
&\hbox{\small $\dot a>0$, $\ddot a<0$}\\ \hline 
\hbox{\small Hubble parameter}
&\hbox{\small $H>0$, $\dot H>0$}
&\hbox{\small $H>0$, $\dot H<0$}\\ \hline
\hbox{\small Event/Particle horizon}
&\hbox{\small Decreasing/--}
&\hbox{\small --/increasing  }\\ \hline 
\hbox{\small 11D Curvature}
&{\small Increasing}
&{\small Decreasing }\\ \hline 
\hbox{\small Initial curvature scale}
&{\small Arbitrarily small}
&{\small Arbitrarily large}\\ \hline 
\hbox{\small 4D effective coupling $g$}
&\hbox{\small $\dot g>0$, $g_i=0$, $g_e=\infty$}
&\hbox{\small $\dot g<0$, $g_i=\infty$, $g_e=0$}\\ \hline
\hbox{\small 4D coupling $g^{(4)}$}
&{\small $\dot g^{(4)}>0$, $g^{(4)}_i=0$}
&{\small $\dot g^{(4)}$ undefined, $g^{(4)}_e=0$}\\ \hline 
\end{tabular}
\bigskip
\end{center}
where $g^{(4)}=e^{\Phi_4}$, $H=\dot a/a$ and a dot denotes differentiation
with respect to cosmic time $t_c$. 

The $(+)$ branch enjoys all the properties of the PRBB branch of the standard
PRBB model \cite{prebb}. So our model leads to a picture of the evolution of
the Universe which does indeed describe a PRBB scenario. According to the
latter, the (four-dimensional) universe starts in an expanding weak-coupled,
low-curvature regime with growing curvature and growing four-dimensional string
coupling $g^{(4)}$. The expansion is accelerated and continues until the
strongly coupled regime with large curvature is approached. Here
non-perturbative effects enter into play and possibly induce a transition to
the POBB $(-)$ branch (graceful exit). The expansion is now decelerated and
both the Hubble parameter and the four-dimensional coupling constant $g^{(4)}$
vanish at large times. 

The consistency of this picture requires that the contribution of the axion to
the Hamiltonian (\ref{II7}) remains subdominant during the PRBB phase. The
axion potential term can be shown to be monotonically decreasing in the region
of the moduli space $\rho>\xi+1/3$. Therefore, a consistent description of the
dynamics constrains the physical solutions to the range $\xi<-17/48\approx
-0.354$ of the $(+)$ branch. This region of the moduli space coincides
essentially with the region $\xi<-1/4$ so an asymptotically past trivial patch
of spacetime with dominant 4-form, which is initially expanding, is likely
to continue its (accelerated) expansion until it reaches the strong-coupling
regime. Therefore, we conclude that a patch of spacetime with dominant
4-form, which is initially expanding and asymptotically past trivial,
potentially evolves to a (spatially flat) homogeneous, accelerated expanding
universe with finite negative deceleration parameter and infinite curvature
(PRBB phase). 

Up to now we have not discussed the dynamical behavior of the internal space
and of the 11th-dimension. The $\xi<-1/4$ region of the $(+)$ branch is
characterized by an expanding internal space for $\xi\le-3/4-1/\sqrt{3}$, and
a bouncing internal space for $-3/4-1/\sqrt{3}<\xi<-1/4$. In both cases the
internal space becomes exponentially large when the strongly coupled region is
approached. The ratio between the two scale factors is 
\be
{a\over R_{C_6}}=e^{\alpha_0-\beta_0}{e^{\tau(\rho-\xi)}\over\sqrt{\cosh(\tau)}}
\,.
\label{III1}
\ee
In the region $\rho>\xi+1/2$ of the moduli space $(\xi,\rho)$ the scale factor
of the internal space grows at a slower pace than the external scale factor.
Therefore for $\xi<-7/12$ the $(+)$ branch solution leads to $a/R_6>>1$ at
large times. Since $\xi<-7/12<-17/48$, the demand that the size of the
six-dimensional internal space is compactified with respect to external space
is consistent with the axion potential remaining subdominant during the PRBB
phase. Now let us turn to the 11th dimension. We have
\be
{a\over R_{S_1}}=e^{-(\alpha_0+\phi_0+6\beta_0)/2}
{e^{2\xi\tau}\over\sqrt{\cosh(\tau)}}\,.
\label{III2}
\ee
In the region $\xi<1/4$ the ratio $a/R_{S_1}$ tend to zero at large times. The
size of the 11th dimension becomes much larger than the size of the
four-dimensional world both for the (expanding) $(+)$ and $(-)$ branches. This
implies that a weakly coupled 11-dimensional universe reaches the
strong-coupling regime in a state with four large dimensions (where matter
exists), one extra-large dimension, and six compactified internal dimensions.
From the point of view of a four-dimensional observer, the universe is
five-dimensional at the end of the PRBB evolutionary phase. This picture shows
some resemblance to the brane world picture of Lukas {\it et al} \cite{LOW},
and of Randall and Sundrum \cite{RS}.

The universe emerges from the graceful exit in a five-dimensional state and
starts evolving according to the $(-)$ branch. During the POBB evolution the
11th dimension continues to expand at a faster pace than the four-dimensional
world. So we expect one extra-large dimension at POBB late times as well.
During the POBB phase the six-dimensional internal space is also expanding
faster than the four-dimensional scale factor. However, this should not be
disturbing because the NS-NS axion potential term will eventually dominate the
R-R 4-form potential. If this occurs, at late POBB times the dynamics is
described by the solution \cite{CM}
\be
\begin{array}{llllll}
\alpha&=&\displaystyle\tilde\alpha_0+{1\over 2}\ln
\left[\cosh\left(\tau\right)\right]
-\tilde\xi\tau\,,\qquad
&p_\alpha&=&\displaystyle 3\tanh\left(\tau\right)-6\tilde\xi\,,\\\\
\phi&=&\displaystyle\tilde\phi_0-{1\over 2}\ln
\left[\cosh\left(\tau\right)\right]+
3\tilde\xi\tau\,,\qquad
&p_\phi&=&\displaystyle\tanh\left(\tau\right)-6\tilde\xi\,,\\\\
\beta&=&\displaystyle\tilde\beta_0+{\tilde p_\beta\over 12}\tau\,,\qquad
&\sigma&=&\displaystyle\tilde\sigma_0+{1\over
\tilde p_\sigma}\tanh\left(\tau\right)\,,
\end{array}
\label{III3}
\ee
where
\ba
1-12\tilde\xi^2+{\tilde p_\beta^2\over 12}-2H&=&0\,,
\label{III4a}\\
3\tilde\alpha_0+\tilde\phi_0-\ln\left|\tilde p_\sigma\right|&=&0\,.
\label{III4b}
\ea
The structure of moduli space of the NS-NS solution is similar to the structure
of the moduli space of the R-R solution. The physical points are determined in
the plane $(\tilde\xi,\tilde p_\beta)$ by the two branches of the hyperbola
(\ref{III4a}) (see figure 3)
\be
\tilde p_\beta=\pm{2\sqrt{3}}\sqrt{12\tilde\xi^2-1}\,.
\label{III5}
\ee
\begin{figure}
\centerline{\epsfxsize=3.6in \epsffile{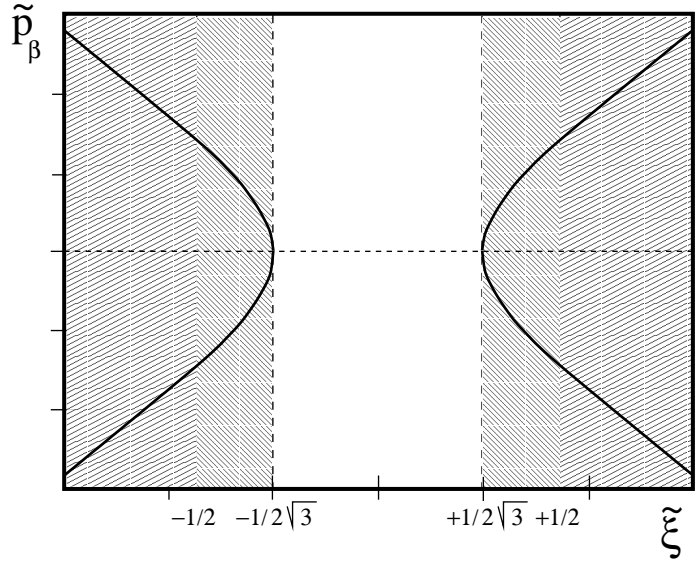}}
\caption{\small Parameter space for the axion-dominated phase. The physical
points are represented by the two branches of the hyperbola (\ref{III5}).}
\label{figIII}
\end{figure}
The left $(-)$ branch and the right $(+)$ branch are characterized by the same
kinematical and dynamical properties of table 1. The dynamical behaviour of the
scale factor of the external space is similar to that of the 4-form
dominated solution. The $(-)$ branch describes a solution with expanding
four-dimensional scale factor for $\tilde\xi\le-1/2$ and bouncing (first
contracting then expanding) four-dimensional scale factor for
$-1/2<\tilde\xi<-1/2\sqrt{3}$. The $(+)$ branch describes bouncing (first
contracting then expanding) and contracting four-dimensional scale factor for
$1/2\sqrt{3}<\tilde\xi<1/2$ and $\tilde\xi\ge 1/2$, respectively.  The sign of
$\tilde p_\beta$ determines the behaviour of the internal space. Positive values
of $\tilde p_\beta$ describe solutions with expanding internal dimensions and
$\tilde p_\beta<0$ solutions with shrinking internal space. The ratio between
the two scale factors is
\be
{a\over R_{C_6}}=e^{\tilde \alpha_0-\tilde \beta_0}
e^{-\tau(\tilde p_\beta/12+\tilde \xi)}\sqrt{\cosh(\tau)}\,.
\label{III6}
\ee
When the axion starts dominating the dynamics of the R-R POBB solution the
11-dimensional universe may find itself either in the $(-)$ or in the $(+)$
branch. Accordingly, the four-dimensional world may either: (a) continue its
decelerated expansion; (b) contract for a while and then resume its (decelerate)
expansion (bouncing solutions); or (c) start an accelerated contracting phase. 
If we live in the axion-dominated phase, a physical expanding universe at large
times requires that the axion-dominated four-dimensional world continues its
expansion in the (a) region of the $(-)$ branch. This is achieved by $\tilde
\xi<-1/2$. From Eq.\ (\ref{III6}) we find that the ratio $a/R_{C_6}$ becomes
exponentially large for the $(-)$ branch solution. As a consequence, the
expansion of the internal dimensions of the R-R POBB phase is eventually
halted. Since the size of the six-dimensional internal space is already
compactified to small scales by the R-R PRBB phase  the internal space remains
compactified at late times.

Now let us turn to the 11th dimension. For the axion-dominated POBB phase
the ratio between the size of the 11th dimension and the scale factor of the
four-dimensional world is
\be
{a\over R_{S_1}}=e^{-(\tilde\alpha_0+\tilde\phi_0+6\tilde\beta_0)/2}
e^{-\tau(\tilde p_\beta/4+\tilde\xi)}\,.
\label{III7}
\ee
The four-dimensional scale factor increases at a faster pace than the
eleven-dimensional radius for the $(-)$ branch. Eventually, for large times we
have $a/R_{S_1}>>1$. However, when the axion starts dominating the dynamics,
the ratio $a/R_{S_1}$ is very small because of the R-R PRBB phase. Therefore,
the universe can be tuned to remain five-dimensional on very long time scales.
\section{Justifying the PRBB-POBB transition}
The M-theory PRBB model which is described in the previous section requires a
transition at high-curvature scales from the $(+)$ branch to the $(-)$ branch.
This transition, the so-called {\it graceful exit} \cite{exit}, is typical of
the PRBB scenario. One of the main unsolved problems of string cosmology is
actually understanding the mechanism responsible for the transition from the
inflationary PRBB phase with increasing curvature to the deflationary POBB
phase with decreasing curvature. Since in the PRBB phase the curvature is
increasing monotonically, the graceful exit necessarily involves a
high-curvature, strongly coupled, regime where higher derivatives and string
loop terms must be taken into account \cite{GMaV}. In the usual PRBB scenario
it has been shown that for any choice of the (local) dilaton potential no
cosmological solutions that connect smoothly the PRBB and POBB phases exist at
classical level \cite{Grex}. In contrast, quantum effects may induce a
transition from the PRBB phase to the POBB phase. A number of quantum string
cosmology models have been investigated in the literature \cite{QSC}. The
outcome of these investigations is that the quantum PRBB-POBB transition
probability is generally finite and non-zero. In the quantum cosmology context,
the transition from the PRBB phase to the POBB phase is described by a
scattering of the PRBB wavefunction by an effective potential barrier that
mimics the strongly coupled regime of the theory \cite{QSC}. In the simplest
models the PRBB and the POBB phases are identified, in the weakly coupled
region of the phase space, by stationary eigenfunctions of the Wheeler-de Witt
(WDW) equation with opposite momentum, say, $\psi^{+}$ and $\psi^{-}$. The
PRBB-POBB transition amplitude is then given by the product
$A=(\psi^{+},\psi^{-})$ in the Hilbert space.

In our model the transition from the $(+)$ branch to the $(-)$ branch may be
explained by a similar mechanism which involves reflection of wavefunctions. 
Let us define the gauge-invariant canonical pairs
\be
\displaystyle
\begin{array}{l}
\displaystyle
X={1\over
5}\left[-6(5\alpha+6\beta+\phi)+{5p_\alpha+3p_\beta-3p_\phi\over\kappa}
\arccosh\left({\kappa\over |{\cal Q}|}e^{-(3\alpha-6\beta-\phi)/2}\right)\right]\,,\\\\
\displaystyle
P_X=-{1\over 48}\left(5p_\alpha+3p_\beta-3p_\phi\right)\,,\\\\
\displaystyle
Y={1\over 5}\left[-12\left(\beta+\phi\right)+{p_\beta-6p_\phi\over\kappa}
\arccosh\left({\kappa\over |{\cal Q}|}e^{-(3\alpha-6\beta-\phi)/2}\right)\right]\,,\\\\
\displaystyle
P_Y={1\over 12}\left(p_\beta-6p_\phi\right)\,,
\end{array}
\label{IV1}
\ee
where
\be
{1\over 16}\left(-p_\alpha+p_\beta-p_\phi\right)^2+
{\cal Q}^2e^{3\alpha-6\beta-\phi}=\kappa^2\,.
\label{IV2}
\ee
The canonical variables above can be completed by the pair $(T,H)$, where
\be
T={1\over\kappa}
\arccosh\left({\kappa\over |{\cal Q}|}
e^{-(3\alpha-6\beta-\phi)/2}\right)\,,
\label{IV3}
\ee
to give a complete set of canonical variables. Using this canonical set the
gauge-fixed density action reads
\be
S_{\rm eff}=\int dt \left[\dot X P_X+\dot Y P_Y -H_{gf}\right]\,,
\label{IV4}
\ee
where we have fixed the gauge $T=t-t_0$. The effective Hamiltonian $H_{gf}$
vanishes on-shell. The Schr\"odinger equation is
\be
\hat H_{gf}\Psi(X,Y;t)=i{\partial~\over\partial t}\Psi(X,Y;t)\,.
\label{IV5}
\ee
Since the effective Hamiltonian of the system is identically zero the wave
functions do not depend on $t$. An orthonormal basis in the Hilbert space is
given by the set of eigenfunctions of the gauge invariant observables,
\be
\hat P_X=-i{\partial~\over\partial X}\,,\qquad
\hat P_Y=-i{\partial~\over\partial Y}\,,
\label{IV6}
\ee
with eigenvalues $x$ and $y$, respectively, 
\be
\Psi(X,Y)={1\over 2\pi}\,e^{i(xX+yY)}\,.
\label{IV7}
\ee
As was expected, in the low-energy limit the wavefunctions are free plane waves
in the two-dimensional $(X,Y)$ space with respect to the Hilbert product
\be
(\psi_1,\psi_2)=\int dXdY~ \psi_1^\star\cdot \psi_2\,.
\label{IV8}
\ee
Quantum high-curvature effects generate a potential $V(X,Y)$. So we expect that
when quantum effects are properly taken into account scattering and reflection
of wavefunctions occurs in the $(X,Y)$ space. Classically, the relation between
the moduli parameters and the gauge-invariant variables is
\be
\begin{array}{lll}
\displaystyle \alpha_0=-{5\over 48}X-{1\over 4}
\ln{|Q|}\,,
&\displaystyle 
\beta_0=-{1\over 16}X+{1\over 12}Y+{1\over 4}
\ln{|Q|}\,,
&\displaystyle 
\phi_0={1\over 16}X-{1\over 2}Y-{1\over 4}
\ln{|Q|}\,,\\\\
\displaystyle
\xi={1\over\kappa}P_X\,,
&\displaystyle
\rho={1\over 5\kappa}(3P_X+P_Y)\,,
&\displaystyle
\chi=-{3\over 5\kappa}(P_X+2P_Y)\,,
\end{array}
\label{IV9}
\ee
where
\be
{12\over 5}(P_Y^2-4P_X^2)=\kappa^2-2H\,.
\label{IV10}
\ee
A reflection in the $(X,Y)$ space with respect to the $X$ plane ($Y\to -Y$) is
equivalent, in the moduli space $(\xi,\rho)$, to the transformation
$\xi\to\xi$, $\rho\to 6\xi/5-\rho$, i.e., to a change of branch in the moduli
space. Therefore, high-curvature quantum effects may induce a transition from
the PRBB branch to the POBB branch, in complete analogy with the standard PRBB
scenario. 
\section{Concluding remarks}
In this paper we have discussed a cosmological scenario which emerges from
(low-energy) M-theory. We have found that the APT postulate and the confinement
of the non-gravitational fields on a 3-brane propagating in the
11-dimensional spacetime lead naturally to a PRBB model. In the latter the
four-dimensional universe undergoes first a phase of accelerated expansion
(PRBB) and then a phase of decelerated expansion (POBB) connected by a state
with large curvature.

Our model is clearly too simple to have any pretense of describing the real
world. A (partial) list of potential problems and drawbacks contains the
following issues: (a) there is no stabilization of the extra-dimensions; (b)
three-dimensional spatial curvature has not been considered; (c) when the full
phase space of the model (4-form + axion) is considered the dynamics is
(probably) non-integrable so at a given evolutionary stage chaotic behaviour
may appear; (d) chaotic behaviour certainly appears when all supergravity form
fields are excited \cite{DH}, so the overall dynamics may be qualitatively
different; (e) fine-tuning issues may appear when trying to fit the observed
cosmological parameters to the model; (f) as in the standard PRBB scenario, we
do not know of any convincing mechanism that could induce the graceful exit,
etc.

Though the list above could include many other potential problems, we still
believe that some interesting information can be extracted from our model.
Firstly, we have learned that the PRBB scenario fits naturally in a model which
is not (at least explicitly) invariant under scale factor duality. This
provides some evidence about the generality of the cosmological PRBB picture in
models derived from M-theory and/or string theory. Scale factor duality may be
an accidental symmetry of the simplest dilaton gravity systems that does not
exist in more complex models\footnote{For instance, $\alpha'$ corrections to
the low-energy string action do not preserve scale factor duality.}, still PRBB
seems to be quite a generic feature. Secondly, R-R forms are recognized to be
essential ingredients in the construction of a viable model of the observed
universe. Finally, and most importantly, we have seen that PRBB and brane-world
cosmologies may be compatible in principle. In this context, it would be worth
trying to implement and reinterpret the PRBB scenario in the LOW \cite{LOW}
and/or RS \cite{RS} brane world models. 
\section*{Acknowledgements}
I am very grateful to Maurizio Gasperini and Paulo Vargas Moniz for interesting
discussions and useful comments. Part of this work was completed during the 8th
International School of Astrophysics Daniel Chalonge. I thank Norma Sanchez for
her kind invitation and the stimulating environment she was able to create in
Erice. I acknowledge many discussions with the participants of the school on
the subject of this paper. I also thank my friends Gianfranco Bertone, Lilith
Haroyan and, in particular, Eun-Joo Ahn for the wonderful time we shared during
the school. 
\end{document}